\documentclass{ws-ijmpcs}

\usepackage{amsmath,amssymb,epsfig,fleqn,url,psboxit,color,here,graphics,graphicx,rotating,multirow,wrapfig}
\usepackage{bigstrut}
\usepackage{hyperref}
\hypersetup{
    colorlinks,
    citecolor=blue,
    filecolor=blue,
    linkcolor=blue,
    urlcolor=blue
}

\newcommand{\slim}{\mskip 1.5mu}              

\newcommand{\phiH}{\phi _h}
\newcommand{\phiS}{\phi _S}

\begin{document}

\markboth{Bakur Parsamyan}{COMPASS-DY: Transverse Spin Physics Program}

%
\catchline{}{}{}{}{}
%

\title{Polarized Drell-Yan at COMPASS-II: Transverse Spin Physics Program}

\author{Bakur Parsamyan}
\address{
ICTP - Strada Costiera 11, 34151 Trieste, Italy
\\
INFN sezione di Trieste - Via Valerio 2, 34127 Trieste, Italy
\\
Universit\`a di Torino \and INFN sezione di Torino -
Via P. Giuria 1, 10125 Torino, Italy
\\
bakur.parsamyan@cern.ch}

\maketitle


\begin{abstract}
Successful realization of polarized Drell-Yan physics program is one of the main goals of the second stage of the COMPASS experiment.
Drell-Yan measurements with high energy (190 GeV/c) pion beam and transversely polarized NH3 target have been initiated by a pilot-run in
the October 2014 and will be followed by 140 days of data taking in 2015.
In the past twelve years COMPASS experiment performed series of SIDIS measurements with high energy muon beam and transversely polarized
deuteron and proton targets. Results obtained for Sivers effect and other target transverse spin dependent and unpolarized azimuthal
asymmetries in SIDIS serve as an important input for general understanding of spin-structure of the nucleon and are being used in
numerous theoretical and phenomenological studies being carried out in the field of transvers-spin physics. Measurement of the
 Sivers and all other azimuthal effects in polarized Drell-Yan at COMPASS will reveal another side of the spin-puzzle providing a
 link between SIDIS and Drell-Yan branches. This will be a unique possibility to test universality and key-features of transverse
 momentum dependent distribution functions (TMD PDFs) using essentially same experimental setup and exploring same kinematic domain.
	In this review main physics aspects of future COMPASS polarized Drell-Yan measurement of azimuthal transverse spin asymmetries
will be presented, giving a particular emphasis on the link with very recent COMPASS results obtained for
SIDIS transverse spin asymmetries from four "Drell-Yan" $Q^2$-ranges.

\keywords{COMPASS; Drell-Yan; SIDIS; Transverse Spin Azimuthal Asymmetries;}
\end{abstract}

\ccode{PACS numbers: 13.60.-r; 13.60.Hb; 13.88.+e; 14.20.Dh; 14.65.-q.}
\section{Introduction}	
\begin{figure}[h]
\centering
\includegraphics[width=12.8cm]{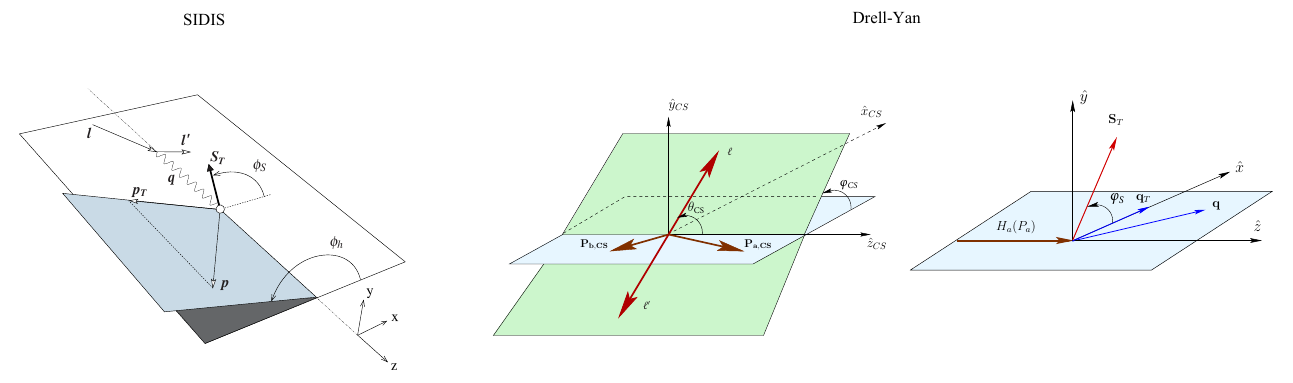}
\caption{SIDIS and Drell-Yan frameworks and notations.. \label{f1}}
\end{figure}

Study of spin-dependent azimuthal asymmetries arising in the SIDIS and Drell-Yan cross-sections is a powerful method used to access TMD distribution functions of the nucleon.
Using standard notations the cross-section expression for the lepton off transversely polarized nucleon SIDIS processes can be written in a following model-independent way\cite{Kotzinian:1994dv}\cdash\cite{Diehl:2005pc}:
{\footnotesize
\begin{eqnarray}\nonumber
&& \hspace*{-1.4cm}\frac{{d\sigma }}{{dxdydzp_{T}^{h}dp_{T}^{h}d{\phiH}d\phiS }} = 2\left[ {\frac{\alpha }{{xy{Q^2}}}\frac{{{y^2}}}{{2\left( {1 - \varepsilon } \right)}}\left( {1 + \frac{{{\gamma ^2}}}{{2x}}} \right)} \right]\left( {{F_{UU,T}} + \varepsilon {F_{UU,L}}} \right) \\ \nonumber
%
%
&&\hspace*{-1.4cm} \times\Bigg\{ 1 + \sqrt {2\varepsilon \left( {1 + \varepsilon } \right)} \textcolor[rgb]{0.00,0.07,1.00}{A_{UU}^{\cos {\phi _h}}}\cos {\phiH} + \varepsilon \textcolor[rgb]{1.00,0.00,0.00}{A_{UU}^{\cos 2{\phi _h}}}\cos \left( {2{\phiH}} \right) + \lambda \sqrt {2\varepsilon \left( {1 - \varepsilon } \right)} \textcolor[rgb]{0.00,0.07,1.00}{A_{LU}^{\sin {\phi _h}}}\sin {\phiH}\\ \nonumber
&&\hspace*{-1.1cm}+\,{{S}_{T}}\Big[\textcolor[rgb]{1.00,0.00,0.00}{A_{UT}^{\sin \left( {{\phiH} - {\phiS}} \right)}}\sin \left( {{\phiH} - {\phiS}} \right) + \varepsilon \textcolor[rgb]{1.00,0.00,0.00}{A_{UT}^{\sin \left( {{\phiH} + {\phiS}} \right)}}\sin \left( {{\phiH} + {\phiS}} \right) + \varepsilon \textcolor[rgb]{1.00,0.00,0.00}{A_{UT}^{\sin \left( {3{\phiH} - {\phiS}} \right)}}\sin \left( {3{\phiH} - {\phiS}} \right)\\ \nonumber
&&\hspace*{+0.1cm}+\,\sqrt {2\varepsilon \left( {1 + \varepsilon } \right)} \textcolor[rgb]{0.00,0.07,1.00}{A_{UT}^{\sin {\phiS}}}\sin {\phiS} + \sqrt {2\varepsilon \left( {1 + \varepsilon } \right)} \textcolor[rgb]{0.00,0.07,1.00}{A_{UT}^{\sin \left( {2{\phiH} - {\phiS}} \right)}}\sin \left( {2{\phiH} - {\phiS}} \right)\Big]\\ \nonumber
&&\hspace*{-1.1cm}+\,{{S}_{T}}\lambda \Big[\sqrt {\left( {1 - {\varepsilon ^2}} \right)} \textcolor[rgb]{1.00,0.00,0.00}{A_{LT}^{\cos \left( {{\phiH} - {\phiS}} \right)}}\cos \left( {{\phiH} - {\phiS}} \right)\\
&&\hspace*{+0.1cm}+\,\sqrt {2\varepsilon \left( {1 - \varepsilon } \right)} \textcolor[rgb]{0.00,0.07,1.00}{A_{LT}^{\cos {\phiS}}}\cos {\phiS} + \sqrt {2\varepsilon \left( {1 - \varepsilon } \right)} \textcolor[rgb]{0.00,0.07,1.00}{A_{LT}^{\cos \left( {2{\phiH} - {\phiS}} \right)}}\cos \left( {2{\phiH} - {\phiS}} \right)\Big]\Bigg\}
\label{eq:SIDIS}
\end{eqnarray}
%
}
where $\varepsilon$ is the ratio of longitudinal and transverse photon fluxes and is given as $\varepsilon = (1-y -\frac{1}{4}\slim \gamma^2 y^2)/(1-y +\frac{1}{2}\slim y^2 +\frac{1}{4}\slim \gamma^2 y^2)$; $\gamma = 2 M x/Q$.
%
%
 Target transverse polarization dependent part of this general expression contains eight azimuthal modulations in the $\phi_h$
and $\phi_S$ azimuthal angles of the produced hadron and of the nucleon spin, correspondingly (see Fig.~\ref{f1}). Each modulation leads to a $A_{BT}^{w_i(\phi_h, \phi_s)}$
Target-Spin-dependent Asymmetry (TSA) defined as a ratio of the associated structure function to the azimuth-independent ones. Here the superscript of the asymmetry
indicates corresponding modulation, the first and the second subscripts - respective ("U"-unpolarized,"L"-longitudinal and
"T"-transverse) polarization of beam and target. Five amplitudes which depend only on $S_T$ are the Single-Spin Asymmetries (SSA), the other three
which depend both on $S_T$ and $\lambda$ (beam longitudinal polarization) are known as Double-Spin Asymmetries (DSA).
Amplitude of each modulation is scaled by a $\varepsilon$-dependent so-called depolarization factor.
Using similar notations, the general form of the single-polarized ($\pi N^\uparrow$) Drell-Yan cross-section (leading order part) in terms of angular variables defined in
Collins-Soper and target rest frames (see Fig.~\ref{f1}) can be written in the following
%
%
model-independent way\cite{Gautheron:2010wva}:
{\footnotesize
\begin{eqnarray}\nonumber     
  &&\hspace*{-1.4cm}\frac{{d{\sigma ^{LO}}}}{{d\Omega }} = \frac{{\alpha _{em}^2}}{{F{q^2}}}F_U^1 \left\{ {1 + {{\cos }^2}\theta  + {{\sin }^2}\theta \textcolor[rgb]{1.00,0.00,0.00}{A_U^{\cos 2{\varphi _{CS}}}}\cos 2{\varphi _{CS}}} \right. \hfill \\ \nonumber
  &&\hspace*{+0.9cm}{\text{   }} + {S_T}\left[ {\left( {1 + {{\cos }^2}\theta } \right)\textcolor[rgb]{1.00,0.00,0.00}{A_T^{\sin {\varphi _S}}}\sin {\varphi _S}} \right. + {\sin ^2}\theta \textcolor[rgb]{1.00,0.00,0.00}{A_T^{\sin \left( {2{\varphi _{CS}} + {\varphi _S}} \right)}}\sin \left( {2{\varphi _{CS}} + {\varphi _S}} \right) \\
  &&\hspace*{+1.9cm}{\text{            }} + \left. {\left. { {{\sin }^2}\theta \textcolor[rgb]{1.00,0.00,0.00}{A_T^{\sin \left( {2{\varphi _{CS}} - {\varphi _S}} \right)}}\sin \left( {2{\varphi _{CS}} - {\varphi _S}} \right)} \right]} \right\}
\label{eq:DY}
\end{eqnarray}
}
Similarly to the SIDIS case, the superscript of the asymmetry indicates the corresponding modulation, while "U","L" and "T" denote the state of the target polarization.
As one can see, in the Drell-Yan cross-section only one unpolarized and three target transverse spin dependent azimuthal modulations arise at leading order.

Within the QCD parton model approach four of
the eight SIDIS TSAs have Leading Order (LO) interpretation (first three SSAs and first DSA in Eq.~\ref{eq:SIDIS})
\footnote{In Eq.~\ref{eq:SIDIS}, Eq.~\ref{eq:DY} and Table.~\ref{tab:PDFs} the LO amplitudes are marked in red and subleading ones - in blue.} and are
described by the different convolutions of Transverse Momentum Dependent (TMD)
twist-two Parton Distribution Functions (PDFs) and Fragmentation Functions(FFs)\cite{Kotzinian:1994dv}\cdash\cite{Mulders:1995dh}
\footnote{In this review FFs in SIDIS and beam PDFs in Drell-Yan are not discussed for brevity.}.
The first two are well-known Sivers (gives access to "Sivers" PDF $f_{1T}^{\perp\,q}$) and Collins (gives access to "transversity" PDF $h_1^q$)
asymmetries\cite{Adolph:2012sn,Adolph:2012sp}. The other two $A_{UT}^{\sin
(3\phi _h -\phi _s )}$ and $A_{LT}^{\cos (\phi _h -\phi _s )}$ LO TSAs are related to the $h_{1T}^{\perp\,q}$ (pretzelosity) and $g_{1T}^q$ (worm-gear) DFs, correspondingly\cite{Parsamyan:2014uda}\cdash\cite{Parsamyan:2007ju,Kotzinian:2006dw,Anselmino:2006yc}).
The remaining four SIDIS asymmetries are suppressed with respect to the leading twist ones by $\sim M/Q$ and are subleading or \emph{higher-twist} effects\cite{Bacchetta:2006tn,Mao:2014aoa,Mao:2014fma}.
However applying wildly adopted so-called "Wandzura-Wilczek approximation" this higher
 twist objects can be simplified to twist-two level (see Refs.~\refcite{Bacchetta:2006tn,Mulders:1995dh} for more details).
The whole set of eight SIDIS asymmetries has been already measured at COMPASS for both deuteron and proton targets
(See Refs.~\refcite{Adolph:2012sn}--\refcite{Parsamyan:2007ju} and references therein).
{\footnotesize
\begin{table}[ph]
\tbl{Nucleon TMD PDFs accessed via SIDIS and Drell-Yan TSAs.}
{\begin{tabular}{@{}ccc@{}} \toprule
  \hline
  SIDIS $\ell^\rightarrow N^\uparrow$ & TMD PDF & DY $\pi N^\uparrow$ (LO)\bigstrut\\ \hline
  \textcolor[rgb]{1.00,0.00,0.00}{$A_{UU}^{\cos 2\phi _h}$}, \textcolor[rgb]{0.00,0.00,1.00}{$A_{UU}^{\cos \phi _h}$} & $h_{1}^{\bot q}$& \textcolor[rgb]{1.00,0.00,0.00}{$A_{U}^{\cos 2\varphi _{CS}}$} \bigstrut\\ \hline
  \textcolor[rgb]{1.00,0.00,0.00}{$A_{UT}^{\sin (\phi _h -\phi _s )}$}, \textcolor[rgb]{0.00,0.00,1.00}{$A_{UT}^{\sin\phi _s}$}, \textcolor[rgb]{0.00,0.00,1.00}{$A_{UT}^{\sin (2\phi _h -\phi _s )}$} & $f_{1T}^{\bot q}$& \textcolor[rgb]{1.00,0.00,0.00}{$A_{T}^{\sin \varphi _{S}}$} \bigstrut\\ \hline
  \textcolor[rgb]{1.00,0.00,0.00}{$A_{UT}^{\sin (\phi _h +\phi _s -\pi)}$}, \textcolor[rgb]{0.00,0.00,1.00}{$A_{UT}^{\sin\phi _s}$} & $h_{1}^{q}$& \textcolor[rgb]{1.00,0.00,0.00}{$A_{T}^{\sin (2\varphi _{CS} -\varphi _{S} )}$} \bigstrut\\ \hline
  \textcolor[rgb]{1.00,0.00,0.00}{$A_{UT}^{\sin (3\phi _h -\phi _s )}$}, \textcolor[rgb]{0.00,0.00,1.00}{$A_{UT}^{\sin (2\phi _h -\phi _s )}$} & $h_{1T}^{\bot q}$& \textcolor[rgb]{1.00,0.00,0.00}{$A_{T}^{\sin (2\varphi _{CS} +\varphi _{S} )}$} \bigstrut\\ \hline
  \textcolor[rgb]{1.00,0.00,0.00}{$A_{LT}^{\cos (\phi _h -\phi _s )}$}, \textcolor[rgb]{0.00,0.00,1.00}{$A_{LT}^{\cos\phi _s}$}, \textcolor[rgb]{0.00,0.00,1.00}{$A_{LT}^{\cos (2\phi _h -\phi _s )}$} & $g_{1T}^{ q}$& \textcolor[rgb]{1.00,0.00,0.00}{double-polarized DY} \bigstrut\\ \hline \\
\vspace*{-20pt}
\label{tab:PDFs}
\end{tabular}}
\end {table}
}
Within the same QCD parton model approach, Drell-Yan TSAs are also interpreted in terms of TMD PDFs.
In this case the asymmetries are related to the convolution of two TMD PDFs: one of the beam and one of the target hadron.
Again, quoting only the target nucleon PDFs: the $A_{T}^{\sin \varphi _s }$, $A_{T}^{\sin
(2\varphi _{CS} -\varphi _s )}$ and $A_{T}^{\sin(2\varphi _{CS} +\varphi _s )}$ give access to the
"Sivers" $f_{1T}^{\perp\,q}$, "transversity" $h_1^q$ and "pretzelosity" $h_{1T}^{\perp\,q}$, distribution functions, respectively.
Within the concept of \textit{generalized universality}\footnote{Time-reversal modified process-independence.} of TMD PDFs it appears that same parton distributions functions can be accessed
both in SIDIS and Drell-Yan (see the Table.~\ref{tab:PDFs} for the complete list).
Therefore, DY measurements at COMPASS will be intriguingly complementary to the COMPASS SIDIS results and
will give an unprecedented opportunity to access TMD PDFs via two mechanisms and test their universality and key features
(for instance, predicted Sivers and Boer-Mulders PDFs sign change) using essentially same experimental setup.
\section{COMPASS: SIDIS -- Drell-Yan "bridge"}	
During it's first phase in 2002-2010 COMPASS has made series of SIDIS TSA measurements using 160 GeV/c
longitudinally polarized muon beam and transversely polarized $^6LiD$ and $NH_3$ targets (See Refs.~\refcite{Adolph:2012sn}--\refcite{Parsamyan:2007ju} and references therein).
New measurements for TSAs, but this time with Drell-Yan reaction are foreseen in 2015 with 190 GeV/c $\pi^-$ beam and transversely polarized polarized $NH_3$-target\cite{Gautheron:2010wva,Quaresma:2014taa}.

\begin{figure}[H]
\centering
\includegraphics[width=12.0cm]{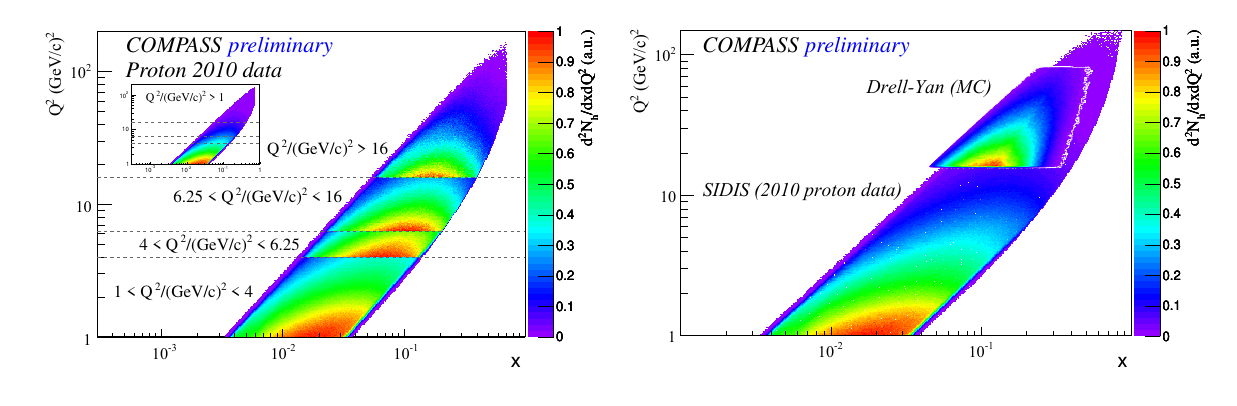}
\caption{COMPASS $x:Q^2$ phase-space with indicated four Drell-Yan $Q^2$-ranges.
\label{fig:f2}}
\end{figure}

Certainly, both sets of COMPASS results from SIDIS and Drell-Yan will become a subject of global fits and phenomenological comparison.
For this purpose the best option is to explore SIDIS data in more differential way extracting the asymmetries in the same four $Q^2$
kinematic regions (which implies also different $x$-coverage) which were selected for the COMPASS Drell-Yan measurement program\cite{Gautheron:2010wva,Quaresma:2014taa}:
{\footnotesize
\begin{eqnarray}
    &&\hspace*{-1.4cm}\bullet\hspace*{0.3cm} 1<Q^{2}/(GeV/c)^2<4\ \ "low\ mass"\\ \nonumber
    &&\hspace*{-1.4cm}\bullet\hspace*{0.3cm} 4<Q^{2}/(GeV/c)^2<6.25\ \ "intermediate\ mass"\\ \nonumber
    &&\hspace*{-1.4cm}\bullet\hspace*{0.3cm} 6.25<Q^{2}/(GeV/c)^2<16\ \ "J/\psi\ mass"\\ \nonumber
    &&\hspace*{-1.4cm}\bullet\hspace*{0.3cm} Q^{2}/(GeV/c)^2>16\ \ "High\ mass". \nonumber
\label{eq:DY_ranges}
\end{eqnarray}
}
In the left plot in Fig.~\ref{fig:f2} COMPASS SIDIS $x$:$Q^2$ kinematical phase-space is shown as divided in four "Drell-Yan" $Q^2$-ranges.
Here the most promising for DY-studies is the so-called "high mass" range which is expected to be free from background and corresponds to the valence-quark region
where the Drell-Yan asymmetries are expected to reach their largest values\cite{Gautheron:2010wva,Quaresma:2014taa}.
On the right plot COMPASS SIDIS $Q^2>1 (GeV/c)^2$ and Drell-Yan "high mass" overlapping distributions are demonstrated.

SIDIS TSAs extracted from all four aforementioned "$Q^2$-ranges"\cite{Parsamyan:2014uda} together with COMPASS recent results for multi-differential analysis\cite{Parsamyan:2015dfa}
will serve not only for future SIDIS-DY comparison,
but, exploring two-dimensional $x$:$Q^2$-behaviour of the asymmetries, they can be used also as a better input
for TMD-evolution studies and related SIDIS-DY predictions \cite{Aybat:2011ta,Echevarria:2014xaa,Sun:2013hua}.

\begin{figure}[H]
\centering
\includegraphics[width=9.0cm]{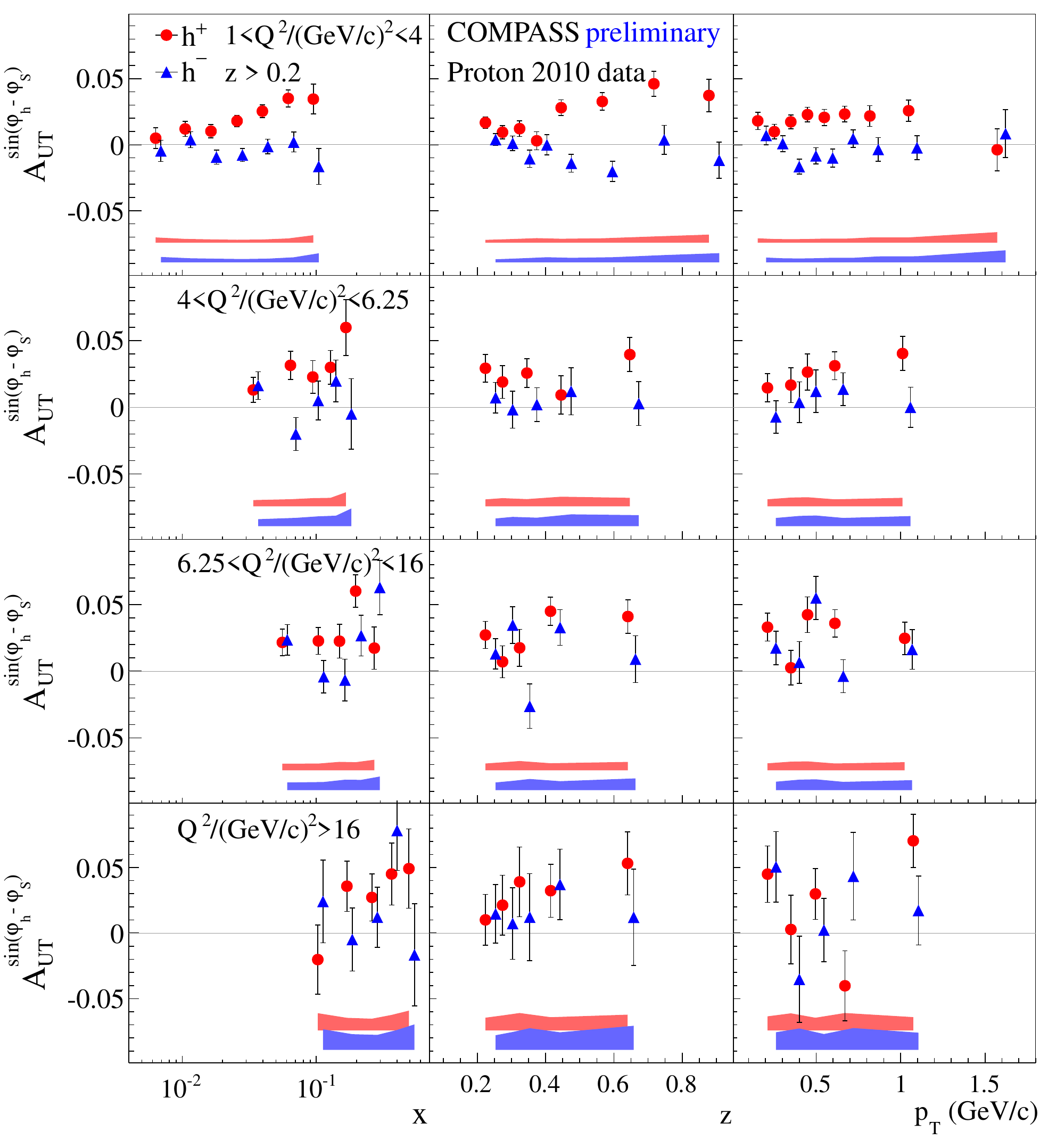}
\caption{SIDIS Sivers asymmetry from Drell-Yan $Q^2$-ranges.}
\label{fig:f3}
\end{figure}

\begin{figure*}[h!]
\centering
\includegraphics[width=3.05cm]{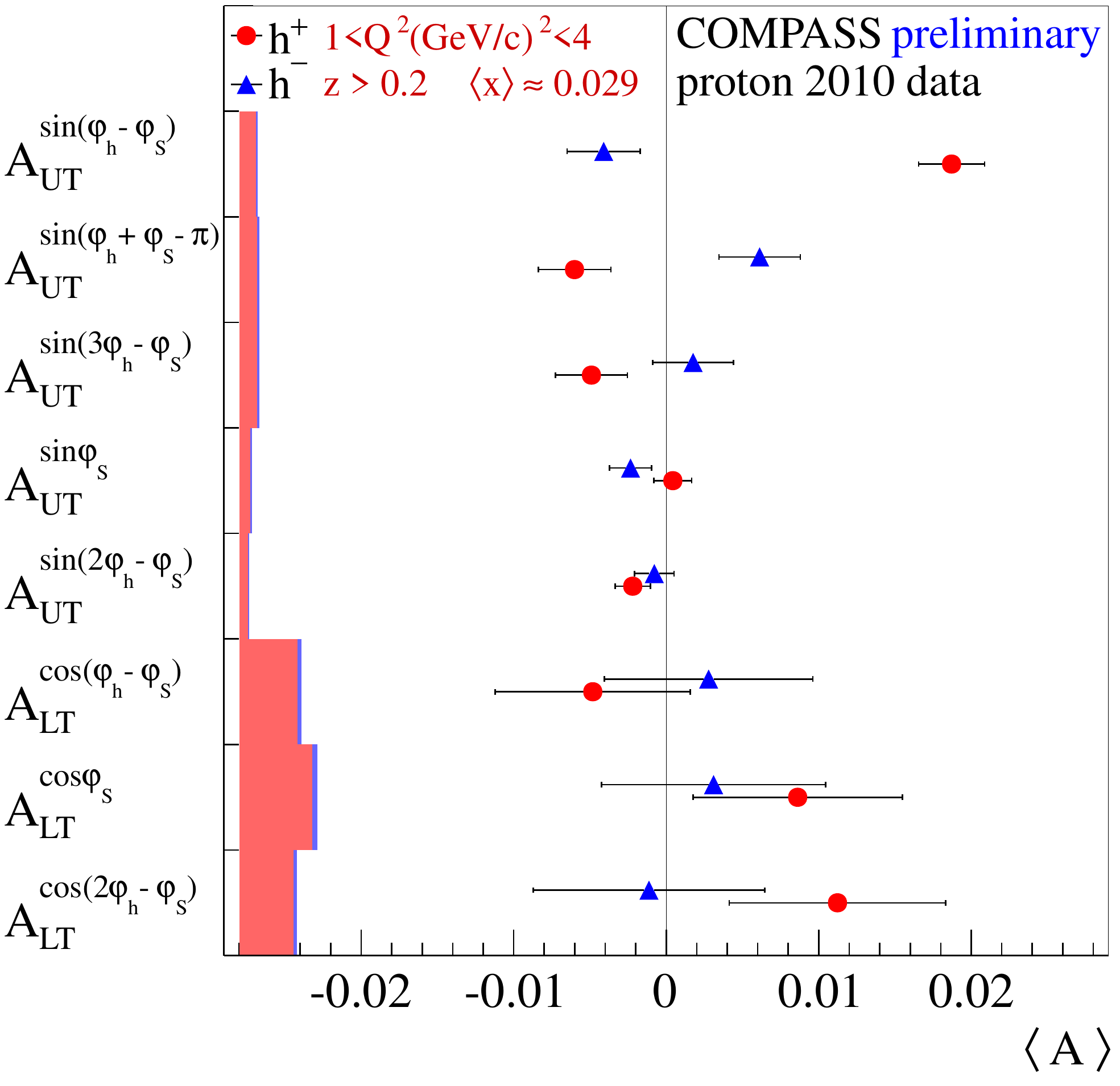}
\includegraphics[width=3.05cm]{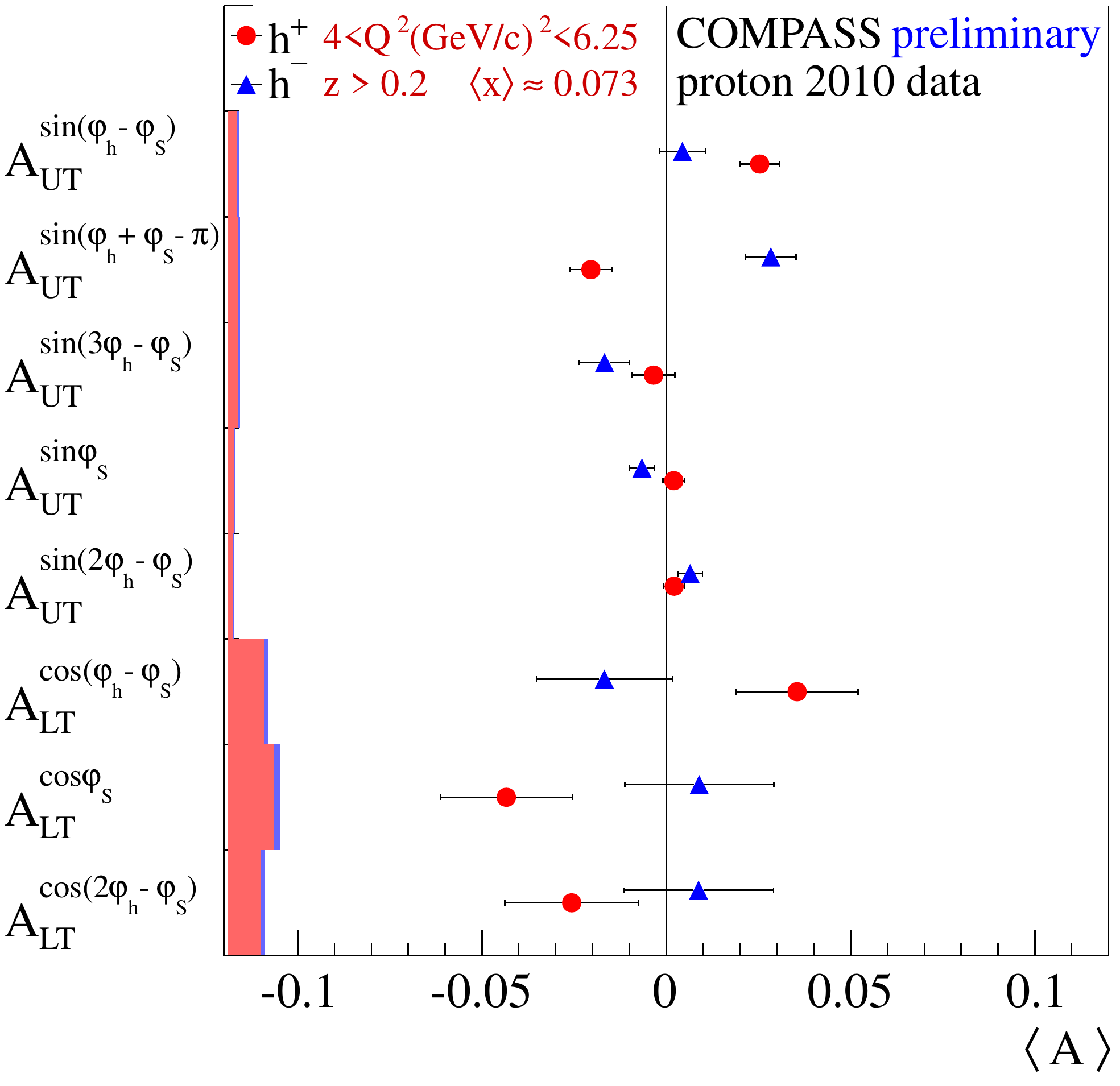}
\includegraphics[width=3.05cm]{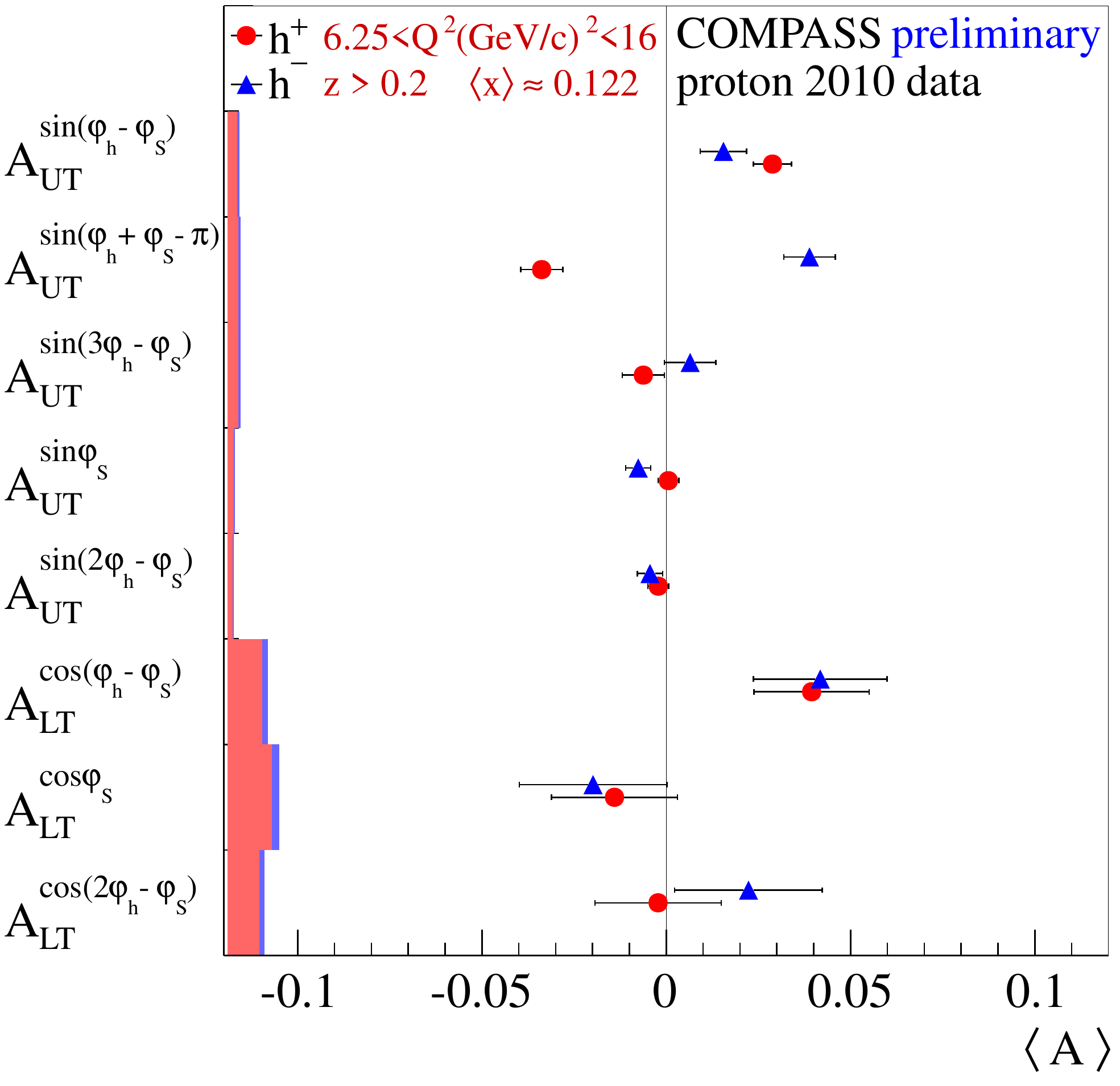}
\includegraphics[width=3.05cm]{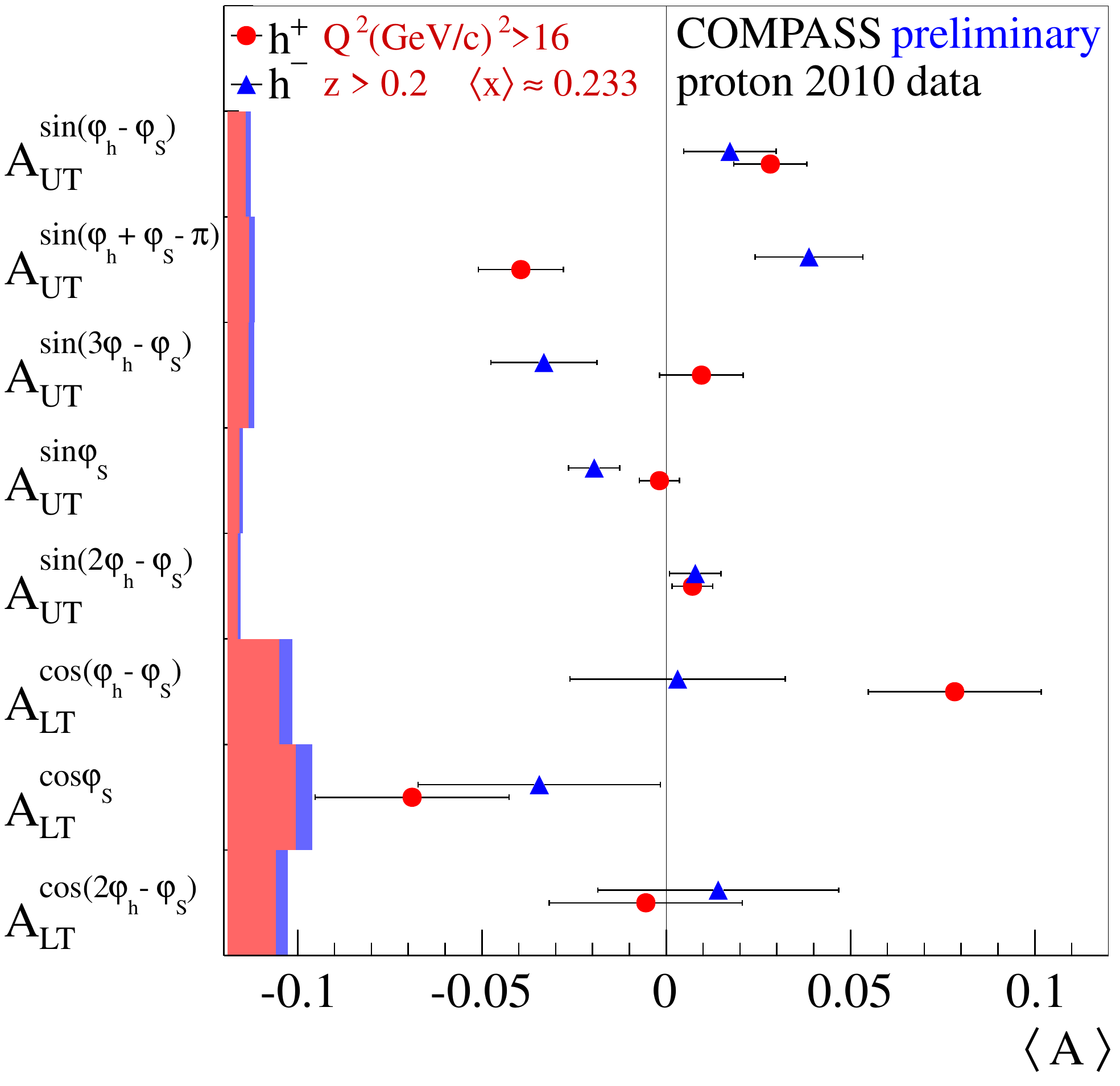}
\caption{Mean SIDIS TSAs in four Drell-Yan $Q^2$-ranges (left to right)}
\label{fig:f4}       
\end{figure*}

Results for Sivers asymmetry are presented in Fig.~\ref{fig:f3}.
A clear positive signal is observed for positive hadrons (growing with $x$, $z$ and $p_T$).
For negative hadrons some hints of a negative amplitude can be seen at lowest $Q^2$-range for intermediate $z$ values
while at relatively large $x$ and $Q^2$ there are indication for a positive signal.
Comparing points from same $x$-bins, but different $Q^2$-ranges one can see that within statistical accuracy there is no clear and strong $Q^2$-dependence for the effect.
Nevertheless, decreasing with $Q^2$ trend can be noted in some bins which has been confirmed also by the recent more detailed multidimensional analysis \cite{Parsamyan:2015dfa}.
It is interesting to mention that for planned Drell-Yan measurements Sivers asymmetry in the "high mass"-range is expected to have approximately
same statistical accuracy as it's SIDIS-analogue extracted from COMPASS proton-2010 data ($\approx1\%$). This should be enough to address the
Sivers function "sign-change" challenge and
possibly also to study kinematical dependences of the effect in several bins \cite{Aybat:2011ta,Echevarria:2014xaa,Sun:2013hua}.
Even if Sivers effect in Drell-Yan is a flagship measurement for COMPASS-II, other TSAs and their links with SIDIS "analogues" from Table.~\ref{tab:PDFs}
are also very important for general TMD PDF-studies. In the Fig.~\ref{fig:f4} mean values for all eight SIDIS-TSAs are quoted as measured in four Drell-Yan $Q^2$-ranges.
\section{Conclusions}
\label{sec:conclusions}
In 2015 COMPASS experiment will start collecting first ever transversely polarized Drell-Yan data.
Recently COMPASS has provided first input for future direct SIDIS-DY studies. Using COMPASS proton 2010 transvers data all eight SIDIS TSAs were extracted from four $Q^2$-ranges
selected for the COMPASS Drell-Yan program.
These results combined with future polarized Drell-Yan data from COMPASS will give a
unique opportunity to access TMD PDFs via two processes and test their universality and key features sticking to the same $x$:$Q^2$ kinematical range.
\\\\
Bakur Parsamyan undertook this work with the support of the ICTP TRIL Programme, Trieste, Italy.
%
%
%
%
%
%
%
%
%

\end{document}